# DESIGNING AN INDUSTRIAL POLICY FOR DEVELOPING COUNTRIES: A NEW APPROACH

ALI HAERI[*]

ABBAS ARABMAZAR[**]

*Abstract:* *In this study, the prevalent methodology for design of the industrial policy in developing countries was critically assessed, and it was shown that the mechanism and content of classical method is fundamentally contradictory to the goals and components of the endogenous growth theories. This study, by proposing a new approach, along settling Schumpeter's economic growth theory as a policy framework, designed the process of entering, analyzing and processing data as the mechanism of the industrial policy in order to provide "theoretical consistency" and "technical and Statistical requirements" for targeting the growth stimulant factor effectively.* (**JEL** *O00, O20, O25, O38, E61, L52*)

[*] M.S in Economics, Shahid Beheshti University, Tehran, Iran. (Corresponding author). Email: al.haeri@mail.sbu.ac.ir

[**] Associate Professor in Economics, Shahid Beheshti University, Tehran, Iran. Email: ab_arabmazar@sbu.ac.ir

[1]

**Introduction**

As countries face resource constraints, designing and implementing the industrial policy to give priorities over industrial branches, become inevitable. Economists usually use individual or composite index for this selection and prioritization, so-called "picking winner". Indices such as "Revealed Comparative Advantages" RCA (Balassa, 1965), "Domestic Resources Cost" DRC (Bruno, 1972), "Growth Poles" index (Perroux, [1955], 1970), "Intensive technology" index (Hatzichronoglou, 1997), "product sophistication" index (Lall, Weiss, Zhang, 2006), "Value-Added Content" index, the index of "Growth Acceleration Of Target Market (demand)" and "Science, Technology and Innovation (STI) indicators" (National Research Council, 2014), used to prioritize industrial groups based on factors such as export advantage, access to raw materials, backward and forward linkages, complexity of production, R&D intensity and patent statistics[1]. Some of these indices emphasize the revealed advantages and continuation of the previous path, while others stress the potential and future competitive advantages of industries.

Data for calculation of each index, are often macroeconomic statistics that are, in most cases, aggregate and each industrial group includes a wide range of non-homogenous industries and companies. The result of processing data is a figure that indicates whether an industrial group should be given the priority in government support or not? Regardless of the inherent constraints of each index which encourages the economist to use a combination of indices; the question is whether it is acceptable to subsume all products and companies of an industrial group under the government support, and, in contrast, disregard all products and companies of another industrial group, by using raw statistics? Even if there are more detailed statistics, since these types of data reflect the current state of the industry and not what they will be in the future, as well as, utilizing them in a combination of heterogeneous indices, would it be possible to adapt the industrial policy to an economic growth theory?

---

[1] There is another type of classification, known as the taxonomy of "sectoral technological trajectories" (Pavitt, 1984) or "technological regimes" (Nelson and Winter, 1983; Winter, 1984; Breschi, Malerba and Orsenigo, 2000; Peneder, 2010) or "technological paradigms" (Dosi, 1982), which put especial emphasize on the importance of the sectoral patterns in technological evolution. Moreover, Cesaratto and Mangano (1993), Arvanitis and Hollenstein (1998), Archibugi (2001) argued for leaving aside "sectoral level" and focus instead on "firm level". For more, refer to Peneder's structuralist study in industry taxonomy: Peneder, Michael. 2003. "Industry Classifications: Aim, Scope and Techniques". Kluwer Academic Publishers, Netherlands. Journal of Industry, Competition and Trade, 3:1/2, 109-129.

Here we propose two approaches: "statistical and technical deepening" and "theoretical deepening" in the form of a new methodology for designing and implementing the vertical industrial policy, to remedy the major damage caused by the classical method. For this purpose, in section 1, while defining precisely the vertical industrial policy, we will refer to the design and implement of industrial policy in industrialized (developed) and developing countries. In section 2, along with the critical assessments of the applied (classical) methods for the design and implement of industrial policy in developing countries, we will discuss the logical and scientific bases of "statistical and technical deepening" and "theoretical deepening", and the necessity of their application as a mean and requirement to eliminate the deficiency of the classical method. In Section 3, a new system for the design and implement of industrial policy is presented to give operational capability to the two mentioned approaches within this system. In section 4, we will mention the policy considerations according to Schumpeter's theory. And in Section 5, we will conclude the discussions.

**1. Literature review and the domain of industrial policy**

In this study, what we mean by "industrial policy", is "a government policy for targeted industries that deliberately favors sectors or industries (or even firms) over others—which is usually (not necessarily) against market signals—to enhance efficiency and promote productivity growth for the whole economy as well as for the targeted industries, which is referred to as selective industrial policy or sectoral industrial policy or vertical industrial policy" (Chang, Andreoni, Kuan, 2013). Industrial policy can also be defined horizontally (functional), which is related to public goods, such as education, R & D, and infrastructure. This is not considered in our definition, because its' domain is not very clear. For example, if a government subsidizes R&D activities, it has actually targeted R&D-intensive hi-tech sectors, and this is a sectoral industrial policy. Hausman and Rodrik (2006) also show that even a specific fiscal, monetary or exchange policy is in the interests of some sectors and industries and to the detriment of others. So even governments that are claiming neutral policies toward industries, are doomed to choose and use priorities for industries.

Based on this fact, many governments have favored selected industrial groups in the form of a vertical industrial policy. In their study, Chang, Andreoni and Kuan (2013), have explored the industrial policy experiences in a wide range of countries, including Japan, United States, Germany, Korea, Singapore, Finland, Italy, Brazil and China, and they show that the method for designing and implementing the industrial roadmap and selecting industrial groups, vary from one country to another. Take what happened in Europe recently. Based on the statistical analysis of the European Union as well as the assessment of the 2008 economic crisis and the "sectoral imbalances accumulation" from 1997 to 2009 that led to the formation of bubbles in non-productive sectors — such as construction and related services, including real estate, renting and business activities — and the simultaneous shrinking of the productive sectors such as manufacturing that was believed to have caused the severity and continuation of the economic crisis of the Union (European Commission, 2011, p35-39), the EU industrial policy was designed. The aim of this policy was to stimulate growth and competitiveness in the manufacturing sector in order to create a European industrial renaissance (re-industrialize Europe) with a view to sustain, modern and innovative economy. In this document, the target point for manufacturing sector is "20 percent" share of the gross domestic product by 2020 from its current share of "15.1 percent". And the strategic industrial sectors included: Chemicals; Automotive; Tourism; Textiles, Defense, Fashion and high end industries, Creative industries, Raw materials, Metals, Minerals and forest-based industries, Mechanical engineering industries, Electrical engineering industries, Food, healthcare, Biotechnology, Aeronautic and Maritime industries. These industrial sectors had to be considered for government support. Additionally, in this document, other sectors with geostrategic implications and a high degree of public intervention, namely, defense, security and space; and high-potential areas including Key Enabling Technologies (KETs) were recognized. KETs comprise micro and nanoelectronics, nanotechnology, industrial biotechnology, advanced materials, photonics and advanced manufacturing technologies. These six technologies provide the basis for innovation in a wide range of industries. Countries and regions, that fully exploit KETs, will be at the forefront of advanced and sustainable economies. (European Commission, 2014).

But what potentially threatens the success of an industrial policy is the difference in design and implement of an industrial policy (the selection of targeted industries) between an industrialized economy and a developing economy. For example, the design and implement of an industrial policy in a developed (industrialized) economy— as seen in the selection of targeted industries and technologies in "High-Tech Strategy 2020 for Germany" (Federal Ministry of Education and Research, 2010) —is more clear, more simple, more dynamic and more reliable, due to German industries catching-up with technological boundaries. In contrast, in developing countries, the more the technology gap, the longer the process of capability accumulation and technological learning in the industry, and the time factor increases the risk of matching this process with competitive boundaries. So, the selecting industries and technologies for governments becomes more complicated, more sensitive and more risky. For example, the targeting of six KETs in a developing country is more risky than in industrialized countries.

Hence, in a common way, these governments try to reduce the time, cost, and risk of catching-up with competitive advantages by employing economic indices to prioritized industries in designing their industrial policy. In the same way, Iran's Ministry of Industry, Trade and Mine in "Industry, Mine, and Trade Strategic Plan" (first version, 2015), compared the industrial activities based on indices such as the share of value-added, labor absorption, exports, market share, comparative advantage and supply chain (as raw materials or final goods) and the level of knowledge and technology, and then, prioritize them in the following order[2]: 1.Manufacture of refined petroleum products 2.Chemicals and chemical products 3.Non-metallic mineral products 4.Motor vehicles, trailers and semi-trailers 5.Mining and quarrying 6.Basic metals 7.Advanced industries 8.Rubber and plastic products 9.Food products and beverages 10.Machinery and equipment n.e.c[3] 11.Textiles 12.Electrical machinery and apparatus n.e.c. However, after a short time (second version, 2017), using the same indices, updating the macroeconomic data and receiving feedback from the private sector and other government institutions, the ministry—based on what is called in this document, "Rolling Wave Planning" for dynamically adapting with the existing resources and opportunities ahead and new methods—revised the arrangement and priorities of industrial activities. These frequent edits and revisions, which are an integral part of the classical method of designing industrial policy, while contradicting the principle of stability in long-term planning, also indicates the insecurity and superficiality of the statistical inputs that go into the policy-making mechanism.

---

[2] The inputs into indices are Iran macroeconomic data which were gathered by Central Bank of Iran accordance with System of National Accounts (SNA) 1993. [Inter-Secretariat Working Group on National Accounts. 1993. System of National Accounts 1993. United Nations Statistical Commission]

[3] not elsewhere classified.

In the classic method, industrial policy is designed in the form of long-term development programs, and the most of inputs used to compute indices of policy design are static macroeconomic data. Consequently, the policy adaptability to the rapid technological and competitive changes in the global market is fundamentally contradictory. To overcome this contradiction (the conflict of the method and the purpose), fundamental change in the design and implement of industrial policy is necessary. In the next section, we will identify the tools and requirements for this change.

**2. Critical assessments of the classical method used in industrial policy design**

The change in the method of designing an industrial policy, needs the tools and requirements that can be found in the critical assessments of the classical method. Hence, we describe how the content of the classical method is a main threat for a proper industrial policy.

*2.1. Critical assessments of statistical and technical data inputs*

The assessments of statistical and technical data used in the classical method is debatable in two aspects:
First, when we utilize macroeconomic sectoral statistics as input data to prioritization mechanism of the industrial groups, implementation of the industrial policy would face the risk of non-differentiation between products and firms of the preferable industrial group (aggregation bias). For instance, if the car manufacturing sector is chosen to be a top priority for government support, regardless of the level of technology (base, key, pacing, emerging[4]) applied in product or the phase of production process (assembly, Spare Parts, Auto Design,

---

[4] Chris Floyd, 1997, by developing Arthur D. little's model (1991), makes a classification of technologies in terms of their competitive impact.

Car platforms[5]), policy executors compulsively support all of them indiscriminately. For another example, if textile manufacturing is not selected as a priority, although one product in this sector has the highest competitive technology or one company has the highest productivity, this one would be compulsively deprived. In order to mend this complication, in most cases, the economist has to manipulate the results based on his "professional intuition" and "sound judgment". It is worth mentioning that, with the exception of "national economic planning" "in no other field of empirical inquiry has so massive and sophisticated a statistical machinery been used with such indifferent results" (Leontief, 1971, p3). "The truth is that quantitative methods for grappling with the enormous volume of empirical data have seldom produced results significantly superior to those achieved by the traditional procedure. In an empirical science, after all, nothing ultimately counts but results. Most economists therefore continue to rely upon their "professional intuition" and "sound judgment" to establish the connection between the facts and the theory of economics" (Leontief, 1986, p3).

Second, most statistics employed in industrial planning do not show the level of technology utilized in industry and the technological gap, which are the most important parts of data for industrial forecasting and planning. One of the most vulnerable points in industrial planning, is the outshining conception of industrial policy to the minds of both the economist and the society. A kind of megalomania or self-sufficiency, so the economist can prescribe curative intervention for the economy only by the prioritization of industrial groups through inferring and analyzing general statistics, without knowledge of the complex relationships and structures and requirements of the production process. This intervention can be paraphrased as "tinker with the economic system".

---

[5] This classification was presented in "1999 Automotive Industries Conference in Tehran" reported by Atieh Bahar Consulting Group, 2003.

"To deepen the foundation of our analytical system, it will be necessary to reach unhesitatingly beyond the limits of the domain of economic phenomena as it has been staked out up to now. The pursuit of a more fundamental understanding of the process of production inevitably leads into the area of engineering sciences. […]Establishment of systematic cooperative relationships across the traditional frontiers now separating economics from these adjoining fields is hampered by the sense of self-sufficiency resulting from what I have already characterized as undue reliance on indirect statistical inference as the principal method of empirical research.[…] This complacent feeling, as I said before, discourages venturesome attempts to widen and to deepen the empirical foundations of economic analysis, particularly those attempts that would involve crossing the conventional lines separating ours from the adjoining fields."(Leontief, 1971, p4-5). "Most of the economic forecasting business develops its projections in such aggregate terms that relevant details pertaining, for example, to anticipated technical change are either disregarded at the outset or get dissipated in the ascent - or should I say descent - from concrete engineering details to the formation of representative indices or broad statistical aggregates. The data-gatherers and model-builders involved in the planning process will have to break down the barrier that separates economists, in particular academic economists, from experts possessing specialized technical knowledge of various fields of production and consumption, as well as of private and public management."(Leontief, 1976, p9).

To break down this barrier, this study proposes "statistical and technical deepening" approach in the form of a new mechanism for the design and implement of industrial policy via the establishment of an institution called "Technology Assessment Center". There are two necessities of such an institution, both in designing stage and in the implementing stage. The first is based on what extricates the industrial policy from the closed circle of non-technical general statistics and provide the entry into the product space and its production technology. Thus the specialized branch of "Technology Assessment Center" for that particular product, extracts standardized information (which is standardized by the economist). Secondly, this information cannot, in principle, be analyzed by statisticians or anyone else other than the engineers of that discipline or product.

*2.2. Theoretical critical assessments of the classical method*

In the classical method, using the combination of indices which is due to the inherent limitations and fundamental defects of each index, can easily confronts the theoretical integrity of industrial policy with threat. On the other hand, economic planning in the form of an industrial policy is essentially seeking to stimulate economic growth. To achieve this aim, it is necessary to choose an economic development theory, where its' endogenous stimulant factor is clearly stated, and also the industrial policy can be designed within its' framework.

"Economic life changes occur partly because of changes in the data, to which it tends to adapt itself. But this is not the only kind of economic change, there is another which is not accounted for by influence on the data from without, but which arises from within the system, and this kind of change is the cause of so many important economic phenomena that it seems worth while to build a theory for it, and, in order to do so, to isolate it [New Combinations$^6$] from all the other factors of change. What we are about to consider is that kind of change arising from within

---

[6] "Technologically as well as economically considered, to produce means to combine the things and forces within our reach. Every method of production signifies some such definite combination. Different methods of production can only be distinguished by the manner of the combination that is either by the objects combined or by the relation between their quantities. Every concrete act of production embodies for us, is for us, such a combination. This concept may be extended even to transportation and so forth, in short to everything that is production in the widest sense. An enterprise as such and even the productive conditions of the whole economic system we shall also regard as "combinations" This concept plays a considerable part in our analysis." (Schumpeter, [1911], 1934, p14)

"To produce other things, or the same things by a different method, means to combine these materials and forces differently. In so far as the "new combination" may in time grow out of the old by continuous adjustment in small steps, there is certainly change, possibly growth, but neither a new phenomenon nor development in our sense. In so far as this is not the case, and the new combinations appear discontinuously, then the phenomenon characterizing development emerges. For reasons of expository convenience, henceforth, we shall only mean the latter case when we speak of new combinations of productive means. Development in our sense is then defined by the carrying out of new combinations." (Ibid, p65-66) the system which so displaces its equilibrium point that the new one cannot be reached from the old one by infinitesimal steps."(Schumpeter, [1911], 1934, p64).

[10]

Thus, the "theoretical deepening" of industrial policy is important in three aspects: First, the lack of distinction and diagnosis of the economic growth stimulator, at best, degrade an industrial policy to the remedial or insurer factor to stimulate the demand side in the form of a Keynesian policy. Moreover, Hanusch and Pyka (2007) in their study "a Roadmap to Comprehensive Neo-Schumpeterian Economics" show that if monetary and financial policies (Finance and Public sector) are not the fosterer and supplier of the novelties [New Combinations] as the core part of neo-Schumpeterian economics, there would be nothing but stagnation in the entrepreneurial activities and bubble explosion in non-innovative sectors.

Therefore, financial and monetary supportive instruments employed in an industrial policy must be defined in the form of the endogenous economic growth theory, so at the right time and right place, would stimulate economic growth. Otherwise, industrial policy will exacerbate and prolong the crisis and depression.

"Depressions are not simply evils, which we might attempt to suppress, but – perhaps undesirable – forms of something which has to be done, namely, adjustment to previous economic change. Most of what would be effective in remedying a depression would be equally effective in preventing this adjustment."(Schumpeter, [1934], 1989, p115)

Second, the over-concentration of an industrial policy on the revealed comparative advantages and charting industrial policy in pursuance of the previous path (path dependence) -which would suppress the adventurous decisions and disregard the potential and future competitive advantages- is one of the pitfalls of a "weak industrial policy[7]" which is not formed within an endogenous economic development theory framework. A weak industrial policy calibrates the industries based on their current state, static advantages and demand side, and basically it remains silent about what the industries will be faced in the future (the supply side). Moreover, economic planning (industrial policy) is fundamentally anticipating and confronting with what will happen in the future.

---

[7] This term was used and defined in (Astogra, Cimoli, Porcile, 2014, p84).

Schumpeter [1923] (1939) identified three internal factors of changes (development) for the economic system that include changes in tastes (utility functions), changes in quantity (or quality) of productive resources (population, stock and savings) and changes in methods of supplying commodities (innovation). He logically and cogently proved that all changes in consumer's tastes are incident to, and brought about by producer's actions, and also excluded savings (capital accumulation) from the fundamental counter lines of his analytic model. He defined innovation (novelties already used in production function) as an independent distinct internal factor which replaces the utility function and disturbs the system equilibrium to develop the whole economy and also savings, capital interest and entrepreneur's profit. In a word "development" is the fruit of innovation, hence Schumpeter accredit merely production sphere (supply side) as the motive engine of development.

The theory of economic growth, where its motive engine is an endogenous factor in the production function (supply side), is far more capable of designing the industrial policy within its framework than the growth theory emphasizing demand side (consumption stimulation), or the classical economic growth theories that are fundamentally silent about the endogenous stimulus and the disturbing the equilibrium. Here the importance of the growth theory chosen to design industrial policy within its framework, is determined.

Third, sinking in statistical foundations in more detail (product and firm) brings us to the microeconomic field and the creation of a logical link between microeconomics and macroeconomics, requires the emergence of a growth theory that has the potential to build a meso-economics over the microeconomics.

"It is the meso level of an economic system in which the decisive structural and qualitative changes take place and can be observed" (Hanusch, Pyka 2007, p1). "In the generic programme, meso is central. […]. To rely in this programme only on micro and macro is rather like having Hamlet without the prince. Schumpeter made the cast complete by laying the foundations and by contributing theoretically to meso" (Dopfer, 2012, p157).

Given that the sector of industry is often regarded as the driving force and the endogenous growth core of the economy, and also because of the highest degree of compliance with the theoretical requirements above (distinct stimulant factor, emphatic supply side and meso-economics) those provided by Schumpeter's economic growth theory, we conduct the "theoretical deepening of industrial policy" approach by settling Schumpeter's theory of economic growth as a policy framework. Considering the comprehensiveness of this theory in distinguishing the internal factor of economic growth in the form of innovations (new combinations) carried out in production process by entrepreneur, and also considering the precise definition of other components of the economic system including wages, capital, profits, credit and interest, and the business cycle in this theory, supportive stimulant instruments and policy considerations can be crystallized into the most coherent way in the industrial policy design.

This field of policy, namely the invention and adaptation of instruments, criteria and indices of a policy based on the theory of economic growth, is essentially the task of an economist. Hence, in this approach, one of the other structural requirements is the "Economic Core" which together with the "Technology Assessment Center" form the two arms of the "Dynamic Center for Design and Implement[8] of the Industrial Policy".

---

[8] "A plan is not a forecast. The whole idea of planning assumes the possibility of choice among alternative feasible scenarios. Feasibility is the key word."(Leontief, 1976, p7). Therefore, the design and the implement must be done dynamically and simultaneously.

## 3. The new approach to design and implement the industrial policy

As mentioned above; the necessity of "statistical and technical deepening approach" leads industrial policy into the product space and its production technology, as well as based on the need for "theoretical deepening approach" with the settlement of Schumpeter's economic growth theory as the framework of industrial policy, in which "new combinations" (innovations carried out in production) have been emphasized as the growth stimulant. In this paper, by considering the scope of the "product" as the convergence point of these two approaches, the industrial policy focus changes from the industrial sectors and related statistics, to the product space. In the following, we outline "the new approach for the design and implement of an industrial policy" based on product evaluation and technological adaptation of the product to the concept of new combinations, taking into account all the components and considerations of the theory of economic growth.

*3.1. The starting point ("Production plan" entry)*

In the classical method, the input data for designing the industrial policy were mostly the general and aggregated statistics. In this study, based on the theoretical and technical requirements, the input data for designing the industrial policy are "product" and "production plan". In other words, first, the data of "production plan" enter the "Technology Assessment Center" and the engineers of the related field evaluate the feasibility and technological dimensions of product and plan to provide the required information (data) for the "Economic (Theoretical) Core" of "Dynamic Center for Design and Implement of the Industrial Policy".[9] Second, the Economic

---

[9] Henceforward, we employ "Dynamic Center" instead of "Dynamic Center for the design and Implement of Industrial Policy" for brevity.

Core determines the type of product, based on the Schumpeter's economic growth model to see whether it is a Schumpeterian product or not. After that, the decision would be made in the form of Schumpeterian entrepreneurial motive instruments to stimulate the growth. However, the product data were sometimes entered the Economic Core and then sent to the Technology Assessment Center for technical assessment, which are reviewed in details in following sections.

*3.2. Product evaluation process*

The Production Plan was first presented by investors to Dynamic Center. At this stage, an industrial policy should determine whether the Production Plan is in line with the economic development. In other words, is the desired product a Schumpeterian product? Or according to Schumpeter, is the "Production Plan" an innovation carried out in production (New Combination)? According to theoretical foundations, new combinations causing the economic development, are defined only in the following cases:

"(1) The introduction of a new good - that is one with which consumers are not yet familiar - or of a new quality of a good.

(2) The introduction of a new method of production, that is one not yet tested by experience in the branch of manufacture concerned, which need by no means be founded upon a discovery scientifically new, and can also exist in a new way of handling a commodity commercially.

(3) The opening of a new market that is a market into which the particular branch of manufacture of the country in question has not previously entered, whether or not this market has existed before. [This category (4) can be classified in categories (1) or (2)]

(4) The conquest of a new source of supply of raw materials or half-manufactured goods, again irrespective of whether this source already exists or whether it has first to be created.

(5) The carrying out of the new organization of any industry, like the creation of a monopoly position (for example through justification) or the breaking up of a monopoly position"(Schumpeter, [1911], 1934, p66)

Thus, to form an industrial policy based on Schumpeter's theory framework, the compatibility of "production plan" to the above factors should be specified. Here, it is necessary to have more clarity on technological demarcation and the product position in the market, to include functional concepts of new combinations and specify the information standards for Technology Assessment Center and Economic Core. Thus, the following classifications (four Groups) called "Schumpeterian new combinations platform" were presented. See Figure 1.

**FIGURE 1-** SCHUMPETERIAN NEW COMBINATIONS PLATFORM
**SOURCE**: AUTHORS

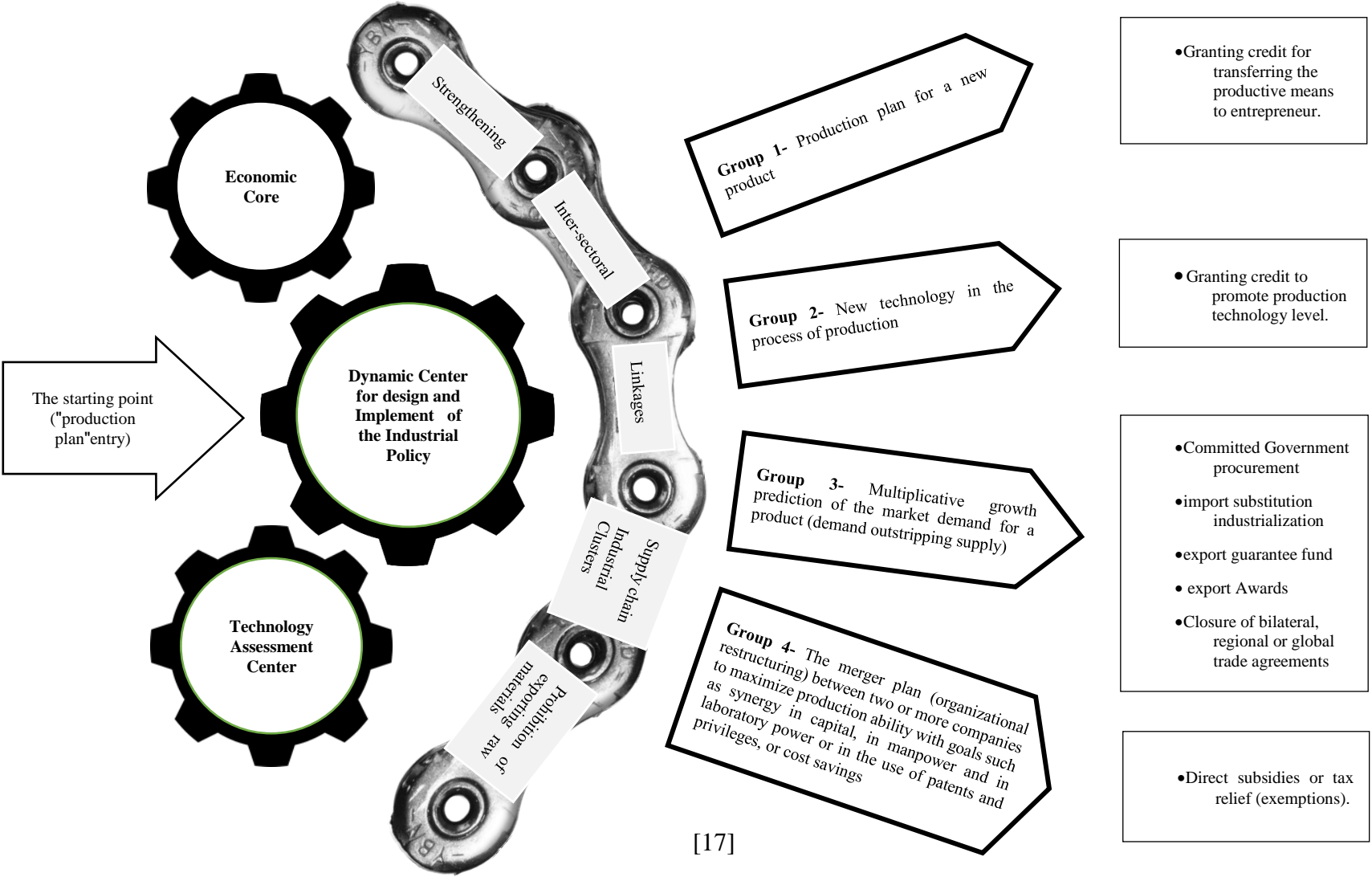

[17]

We explain these four groups as follows:

(Group 1)- <u>The presented production plan to Dynamic Center is a new product which has not presented at least in the internal market.</u> According to Schumpeter, new product is one which replaces a new utility function with the current utility function and changes the taste of consumer. At first, the branch of Technology Assessment Center which is specialized on that product, should confirm that the product has the technical and engineering ability for satisfying an uncovered need of consumer. Next, Economic Core should confirm the probable change caused to utility function and existence of demand for mass production (two words of "probable" and "mass production" are regarded as the keywords). The word "probable" is due to the inherent risk in success of product in the market which always exists and cannot removed. The term "mass production" is raised based on the success of product which should be measured at the stage of mass supply in competitive price on the stock market. So, in these two areas, the economist have no ability of certain judgment. At last, the motive tool suggested by Schumpeterian economy for supporting the mentioned production plan, was the transfer of productive means (capital) to the entrepreneur. This support can be achieved through granting credit (creation of purchasing power by banks) and extending the means of payment in the hands of entrepreneur. Banks must accept the productive means, or products that are supposed to be produced in the future, as collateral.

(Group 2)- <u>The presented production plan to the Dynamic Center provides a new method (technology) in the process of production.</u> In this area, most operations are conducted by Technology Assessment Center. At this stage, the specialized branch of Technology Assessment Center measures the technology level of suggested production method based on the "Technology Content Added" TCA[10] (Asian and Pacific Center for Transfer of Technology, 1987) and specifies the level of technology utilized in the process of production based on "competitive technology classification" (Arthure D. Little Inc, 1991; Floyd, 1997) , including base technologies, key technologies, pacing technologies, and emerging technologies[11]. Since the innovation (the primary wave of Schumpeterian business cycle) and its dispersion to other industries (the secondary wave) increase as we get closer to the last generation of technologies, the base, key, pacing, and emerging technologies respectively have the higher degree of convergence to Schumpeterian economic growth model (see Figure 2). Economic Core should first evaluate the production plan based on the amount of changes caused by the new production process to the TCA. Then, it should prioritize the production plan based on four classes of technology which the production plan would be subsumed one of them. If the technology of production plan includes the first two types, namely the base technology or key technology, the Economic Core should be sensitive to the economic behoof resulted from the reduction of final price in the country. However, if the technology of production plan includes the last two types, namely the pacing technology or emerging technology, the economist have no certain judgment on the economic behoof at the current time because the pacing and emerging technologies are related to the future and many related factors such as unreal wages, low cost of natural resources, variable costs of research and development, are the real limitations faced by economists to determine the behoof of plan in the future. Thus, the last two technologies should be supported in any case.

---

[10] How to measure Technology Content Added (TCA) -based on The Methodology of The Technology Atlas Team presented by Asian and Pacific Center for Transfer of Technology (APCTT), 1987 - is elaborated in Online Appendix A.

[11] The details of this classification are given in the Online Appendix B.

Foreign investments can gain the support of industrial policy in the host country only if the technology utilized in their production process includes the last two types, namely the pacing technology and emerging technology.

(Group 3)- <u>The delivery of production plan to the Dynamic Center was based on the prediction that "demand outstrips supply" for product in the future.</u> The term "prediction" is a risk element for industrial policy. The government cannot invest from public resources based on a prediction, but Schumpeter considered the opening of new markets as the cause of endogenous growth of the economic system. Thus, the risk should be accepted for achieving the benefits resulted from the market multiplicative growth for some products, but the three following methods can minimize the risk.

- The government has to purchase some products based on its future strategic plans in form of political, social, cultural, military, economic and administrative procurement. If the government is committed to its plans, the growth of market demand for the products required by the government is not a prediction but a commitment. Directly, the government can support the production plans of the required products in the future. However, this contract has some provisions such as the Technology Assessment Center should evaluate the quality and quantity of the production plan to match and satisfy the government's need. meanwhile, the Economic Core should prevent the shortage and excess (lumpy) investment in producing and importing the products required by the government in the future. The supportive tools in this method is granting the bank credit for transferring the productive means to entrepreneur with the minimum interest, and imposing the tariffs on imported goods until the domestic industries reach a competitive level.

- If the production plan is based on the multiplicative growth prediction of the internal market and if these predications are made with the best economic tools, it still has the risk of success. But neglecting the growth of internal market in the future and granting this share to foreign products is not be acceptable. In this case, the government should support the production plan in the form of a mutual contract. The term "contract" is a keyword. First, the Technology Assessment Center determines the qualitative standard for estimating and covering the internal demand and the Economic Core controls the shortage or excess (lumpy) of investment in producing and importing the product and provides an opportunity for the convergence of the domestic price to the global (competitive) price. In addition, the supportive tool of the government for this type of product is restricting import by imposing tariffs which is called "import substitution industrialization". This type of protectionist policy is also known as "the government support of infant industries".
- If the production plan is based on the multiplicative growth prediction of the global market, the prediction would get more risky and the situation would get out of government control. In this case, the government can support the production plan in form of some commitments such as export subsidy or export guarantee fund as well as signing the bilateral agreements between the countries and regions to create a new market for manufactured products in the country.

(Group 4)- <u>The merger plan (organizational restructuring) between two or more companies to maximize production power _with goals such as synergy in capital, in manpower and in laboratory power or in the use of patents and privileges, or cost savings_ presented to the Dynamic Center.</u> In this case, the Technology Assessment Center undertake the plan to verify the claimed objectives. Also, the Economic Core should evaluate the merger conditions and validate the merger plan usefulness for the economy by codifying a local standard system[12] to overwatch the competitive conditions (market concentration index) and banning the monopolistic practices and such pitfalls. Taking into account competitive conditions, this merger or organizational restructuring causes the endogenous growth of the economic system based on the Schumpeter's theory. The supportive instruments can be the direct subsidies and tax relief.

*3.3. Strengthening the inter-sectoral linkages*

So far, this study focused on the stimulation of economic growth by utilizing innovation (technology) in product or creation of new markets for manufactured products. However, creating an economic growth will be an incomplete process without providing the context for the continuity and dispersion of economic boom between all economic sectors. According to Schumpeter's business cycle theory, the inter-sectoral forward and backward linkage, both in terms of strength and interlacement, and in terms of the diversity and multiplicity, should be in a situation where the boom resulted by industrial policy (primary wave) is immediately reinforced by inter-sectoral linkages, so this boom would become a sustainable and long-term prosperity in the form of supply growth or demand growth, or the encouragement and spread of innovation to other sectors (secondary wave). See Figure 2.

Eslava, Fieler and Yi Xu (2015) provided evidence on "magnification effect of technology adoption between advanced and backward firms" in their empirical studty. "When a subset of firms adopts newer technologies and managerial practices, they become more stringent in their input purchases and may prod their suppliers to also adopt newer technologies. With economies of scale, the cost of these advanced-technology inputs decreases, which in turn, increases the incentives for other firms that use these same inputs to upgrade their own technology. Analogous spillovers hold for downstream sectors. Firms that adopt newer technologies increase the availability of better inputs and thereby lower their customers' cost of using newer technologies. In other words, the adoption of advanced technologies by a subset of firms may trigger broad improvements in a wide range of firms" (Eslava, Fieler, Yi Xu, 2015, p662). As there are more inter-sectoral relationships (trade-off), the length of the boom (secondary wave) will be greater and the domestic economy will have a larger share of the growth benefits than foreign countries (exporting countries).

---

[12] For example, US Department of Justice and the Federal Trade Commission and (2010), using the Herfindahl-Hirschman Index, have developed specific standards for preventing the formation of trusts and monopolies through mergers and institutional restructures, as detailed in Online Appendix C.

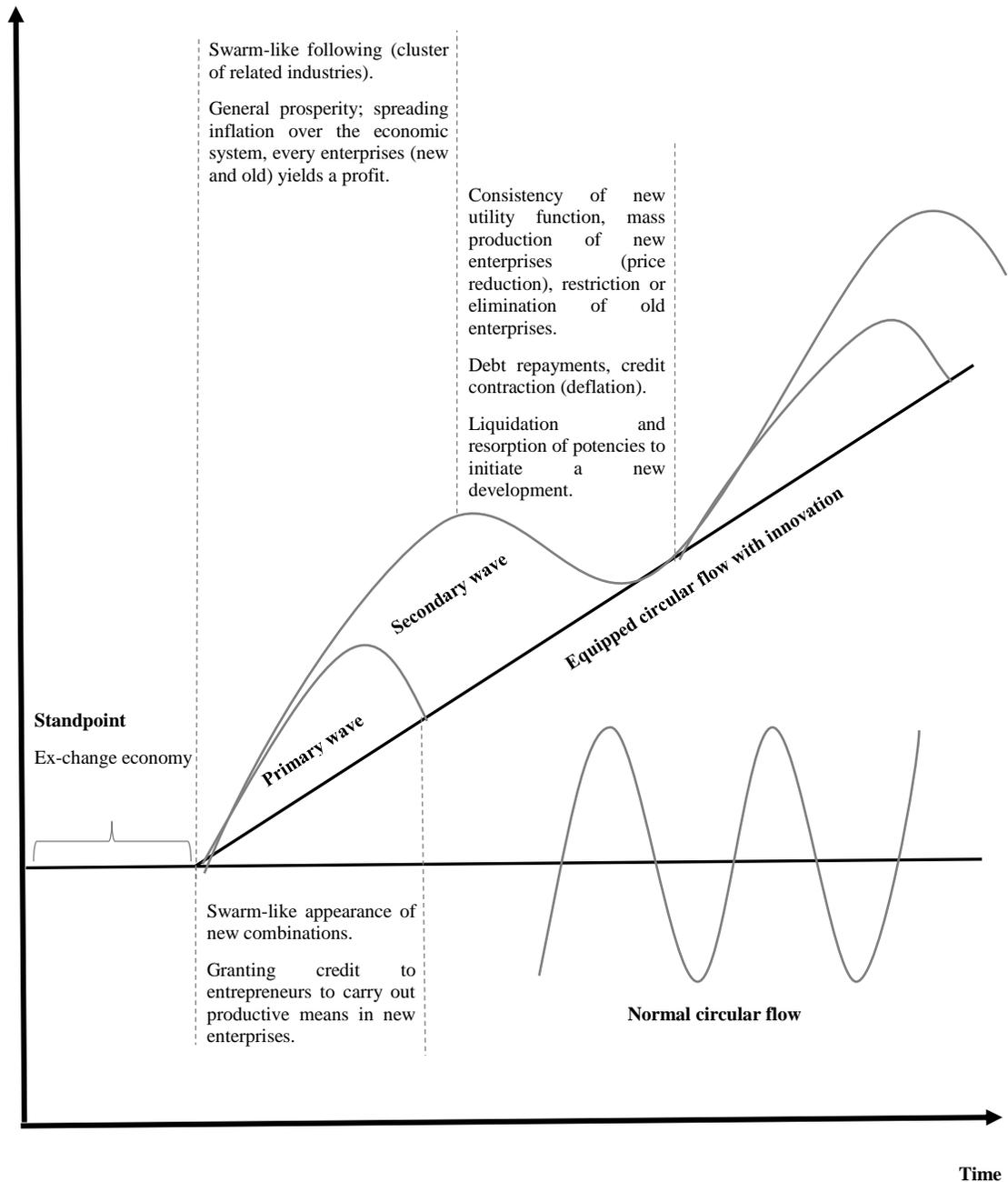

**FIGURE 2-** SCHUMPETERIAN BUSINESS CYCLE
**SOURCE**: AUTHORS

Therefore, the Economic Core -while calculating the backward and forward linkage indices as well as the power and sensitivity of dispersion indices[13] - must overwatch the cells of the input-output table to have a balanced growth. For this purpose, the best strategy is import substitution industrialization, so the production plan of the material (input) -that is currently supplied by import- should be protected by prioritizing the production of intermediate inputs of the supply chain (industrial clusters strategy), and by focusing on the products which use the internal inputs (the prohibition of exporting raw materials strategy)[14]. Accordingly, the production plan is first sent to the Technology Assessment Center to evaluate the technological, qualitative, and quantitative ability of plan in satisfying the internal market demand, compared to the imported products. Then, the Economic Core, based on a contract and taking into account the limited time to support infant industry and the competitive course of the price of the product during the contract, impose protective tariffs on import of the similar product. Because the imposition of restrictions on the import of foreign products is also a kind of creation a new market in the country, the Dynamic Center should provide the conditions to finance the plan by creation of the purchasing power in the hands of the entrepreneur to possess the productive means.

All the above-mentioned processes are based on the Schumpeter's economic growth theory which deepen the industrial policy theoretically, and integrating the industrial roadmap of the country. In addition, deepening the statistical foundations was achieved by settling the Technology Assessment Center, so that the Economic

---

[13] The calculation methods of these indices are presented in Online Appendix D based on the Soofi's work (1992)

[14] But we shouldn't interpret "import substitution" or "prohibition of exporting raw material" as an easy imperative like imposing tariff alone. Kee and Tang (2016) show that China within a purposeful planed policy, has used trade and foreign direct investment (FDI) liberalization since the early 2000s as an opportunity to gradual substitution of domestic for imported materials by its exporters. China trough the continuous tariff reduction facing upstream firms, led FDI to targeted downstream firms. This policy forced upstreaming firms to expand domestic input variety with more competitive price. "China also highlight that trade and FDI liberalization may actually raise a country's domestic value added in exports to gross exports (DVAR), through input-output linkages and spillovers that go beyond the targeted industries" (Kee, Tang, 2016, p1434).

Core of Dynamic Center can design and reform the policy using more realistic, more disaggregated (even micro) and more specialized data.

*3.4. Dynamic Center's supportive instruments*

One of the most important supportive instruments of an industrial policy is "transferring the productive means to entrepreneur". The capital (purchasing power required by entrepreneur) is mainly provided by banks which with the power of creating money[15], act no longer as intermediaries, but as capitalists, So "the only man, he [entrepreneur] has to convince or to impress is the banker who is to finance him" (Schumpeter, [1911], 1934, p89). In this position, the pivotal task of the industrial policy is "to reduce the risk of entrepreneur to the lowest level and transfer the risk to the bank" because "the entrepreneur is never the risk bearer". (Ibid, p137). Here, the role of Development Banks becomes important in industrial planning, so that "interest" should only be defined as a slice of entrepreneur's profit, as well as a cover for the risk of capital (debt) repayment. In other words, any tact that reduces entrepreneurial risk is a supportive instrument of the industrial policy. Granting credit to entrepreneur and taking "productive means" or "production plan" or even "future (coming) products of the plan" as collaterals, transfer the risk of production to banks. Also the public investments, insurance funds, investment guarantee fund, government commitment to purchase commodities in accordance with the contract, imposition of tariffs on similar imports within a limited time and based on a specified contract between the government and the entrepreneur, Closure of bilateral, regional or global trade agreements to facilitate the export of entrepreneur's products, export guarantee fund , export Awards, direct subsidies to innovative industries or indirect subsidies such as tax or excise exemptions, include industrial policy supportive instruments which helps to reduce the risk of entrepreneur.

---

[15] Granting credit by banks to an entrepreneur in the Schumpeterian economy no longer means transferring purchasing power from one person to another, But at first necessarily from the Standpoint, the cash flow moves (creation of money) and then by producing commodities against the same amount of money -Of course, with more total utility – the economic growth is achieved.

## 4. Policy considerations

In this section, some of Schumpeterian components and concepts of economic growth are referred., Growth stimulus factor, entrepreneurship, recession, capital and interest are components which have unbreakable link with industrial policy, and adherence to the theoretical requirements of these concepts makes industrial policy more coherent and rectilinear.

*Growth stimulant factor:* According to the fact that the isolation of growth factor (new combinations) is considered as the basis of Schumpeter's economic growth theory, and a systematic retelling of Friedrich List's sentence (1856) "the power of creating wealth is then vastly more important than wealth itself", so It is necessary to avoid the deviation of industrial policy from the real productive sector of the economy (the range of new combinations) to non-productive sectors that are overwhelmed by economic bubbles, speculations and windfalls.

*Entrepreneurship:* Schumpeter has an exact and meticulous endeavor in defining entrepreneurship, and since this definition has an unbreakable link with his theory, the limitations and requirements of this definition must be considered in the design and implement of industrial policy. In his view, the entrepreneur necessarily does not belong to any social classes, including inventor, capitalist, landowner, robber, benefactor to humanity, genius, worker, engineer or shareholder …[16] . In his definition merely "entrepreneur is anyone seizes a gain when it is immediately before his eyes." (Schumpeter, [1911], 1934, p214). Even if a person was an entrepreneur in the past, this would not essentially entail his future entrepreneurial activities, as Schumpeter say: "the precariousness of the economic position both of the individual entrepreneur and of entrepreneurs as a group, and the fact that when his economic success raises him socially he has no cultural tradition or attitude to fall back upon, but moves about in society as an upstart, whose ways are readily laughed at, and we shall understand why this type has never been popular, and why even scientific critique often makes short work of it"(Ibid, p90) . Therefore, if industrial policy focuses on special few or certain group even who have previously been engaged in entrepreneurial activities in industries, the country will lose a lot of entrepreneurial opportunities. The economic and technological assessment of the production plan is the only thing that industrial policy must focus on.

---

[16] "Because being an entrepreneur is not a profession and as a rule not a lasting condition, entrepreneurs do not form a social class in the technical sense, as, for example, landowners or capitalists or workmen do. Of course the entrepreneurial function will lead to certain class positions for the successful entrepreneur and his family. It can also put its stamp on an epoch of social history, can form a style of life, or systems of moral and aesthetic values; but in itself it signifies a class position no more than it presupposes one. And the class position which may be attained is not as such an entrepreneurial position, but is characterized as landowning or capitalist, according to how the proceeds of the enterprise are used. Inheritance of the pecuniary result and of personal qualities may then both keep up this position for more than one generation and make

Schumpeter also refers to resistances that are against the entrepreneur, so The Dynamic Centre should evaluate the production plan independently of the interest groups and apart from the political, social and consuming norms.

*Recession:* In most cases, the recession has occurred due to the backwardness of the technology level of production or management decay. Therefore, support for belated industries and retarded management will not only be an effectless intervention in the cycle, but also lead to the lack of growth and prosperity of advanced technologies and novelty. Established Industries will only be supported in the event of production modernization and organizational reconstruction.

*Capital and interest:* "Capitol is nothing but the lever by which the entrepreneur subjects to his control the concrete goods which he needs" (Ibid, p116). "We shall define capital, then, as that sum of means of payment which is available at any further enterprise easier for descendants, but the function of the entrepreneur itself cannot be inherited, as is shown well enough by the history of manufacturing families"(Schumpeter, [1911], 1934, p78-79) moment for transference to entrepreneur." (Ibid, p122). So to expand and transform the assets, resources and labor of a country into real capital, the following are recommended:

- Promote the legal status of payment instruments such as cheque, promissory note, future money order in developing countries.
- Development and diversification of bank collateral, and defining the legal status of these collateral in the country's financial system. For example, convert the legal status of professional certificates, business licenses, patents and royalties, production plans, and even future (coming) products of the plan into a class of assets and creating the legal platforms and conditions to mortgage these assets to get credit. In less developed countries, issuing title of deed for lands and the same fixed assets are very helpful in increasing capital available to entrepreneurs.
- Establishment of a mechanism to calculate investments risk in order to assure loans and debts of investors for removing the restrictions on collateral. The creation of venture capital funds as well as the investment guarantee fund are methods that transfer capital to an entrepreneur with a low risk.

## 5. Conclusion

In this paper, two approaches of "theoretical deepening" and "statistical and technical deepening" were emphasized for designing and implementing an industrial policy. In this regard, one pragmatic platform was provided to execute both approaches by means of the maximum adaptation of Schumpeter's theory of economic growth with the purpose of an industrial policy for finding an endogenous stimulant factor of economic system (New Combinations) and supporting and canalizing it to the economic growth. Also an alternative to the classic method, in which the data used to calculate the indices for designing and implementing the industrial policy is "production plan".

[29]

In the new method, the process of analyzing the "production plan" (input data) in two analytical Institutions including "Economic Core" and "Technology Assessment Center", was designed in a way that the exploratory information and standards would be provided for accurately recognizing the stimulant factor of economic growth (new combinations). on the other hand the monetary and fiscal policies as supportive instruments of an industrial policy were determined in complete adaptation with the components of Schumpeter's theory of economic growth, i.e. entrepreneurship and the effective factors in swarm-like appearance of entrepreneurs, money and credit creation, capital, business cycles, profit and interest. So the whole method of the design and implement of industrial policy maintains its consistency within an economic development theory, and the industrial roadmap would be clearly marked.

## REFRENCES

**Archibugi, Daniele**. 2001. "Pavitt'S Taxonomy Sixteen Years On: A Review Article." Economics of Innovation and New Technology 10 (5): 415–25. https://doi.org/10.1080/10438590100000016.

**Arthur D. Little, Inc.**, philip A Roussel, Kamal N Saad, and Tamara J Erickson. 1991. Third Generation R & D: Managing the Link to Corporate Strategy. Massachusetts: Harvard Business School Press. https://archive.org/details/thirdgenerationr00rous.

**Arvanitis, Spyros, and Heinz Hollenstein**. 1998. "Innovative Activity and Firm Characteristics - A Cluster Analysis with Firm-Level Data of Swiss Manufacturing." In 25th Annual Conference of the European Association for Research in Industrial Economics. https://www.researchgate.net/publication/241621900_Innovative_Activity_and_Firm_Characteristics_-_A_Cluster_Analysis_with_Firm-level_Data_of_Swiss_Manufacturing.

**Asian and Pacific Center for Transfer of Technology (APCTT).** 1987. "The Technology Atlas Project." Technological Forecasting and Social Change journal, Volume 32, Issue 1, p1-109. https://doi.org/10.1016/0040-1625(87)90003-5 https://doi.org/10.1016/0040-1625(87)90004-7 https://doi.org/10.1016/0040-1625(87)90005-9 https://doi.org/10.1016/0040-1625(87)90006-0 https://doi.org/10.1016/0040-1625(87)90007-2 https://doi.org/10.1016/0040-1625(87)90008-4

**Astorga, Rodrigo, Mario Cimoli, and Gabriel Porcile.** 2014. "The Role of Industrial and Exchange Rate Policies in Promoting Structural Change, Productivity and Employment." In TRANSFORMING ECONOMIES: Making Industrial Policy Work for Growth, Jobs and Development, edited by José Manuel Salazar-Xirinachs, Irmgard Nübler, and Richard Kozul-Wright, pp84. Geneva: International Labour Office ILO. https://www.ilo.org/global/publications/ilo-bookstore/order-online/books/WCMS_242878/lang--en/index.htm.

**Atieh Bahar Consulting Group.** 2003. "1999 Automotive Industries Conference in Tehran." 2003. https://web.archive.org/web/20080617154335/http://www.atiehbahar.com/Resources/Automotive.htm.

**Balassa, Bela.** 1965. "Trade Liberalisation and 'Revealed' Comparative Advantage." The Manchester School 33 (2): 99–123. https://doi.org/10.1111/j.1467-9957.1965.tb00050.x.

**Breschi, Stefano, Franco Malerba, and Luigi Orsenigo.** 2000. "Technological Regimes and Schumpeterian Patterns of Innovation." *The Economic Journal* 110

(463): 388–410. https://doi.org/10.1111/1468-0297.00530.

**Bruno, Michael.** 1972. "Domestic Resource Costs and Effective Protection: Clarification and Synthesis." Journal of Political Economy 80 (1): 16–33. http://www.jstor.org/stable/1830128.

**Cesaratto, Sergio, and Sandro Mangano.** 1993. "Technological Profiles And Economic Performance In The Italian Manufacturing Sector ∗." Economics of Innovation and New Technology 2 (3): 237–56. https://doi.org/10.1080/10438599300000005.

**Chang, Ha-Joon, Antonio Andreoni, and Ming Leong Kuan.** 2013. "International Industrial Policy Experiences and the Lessons for the UK." Future of Manufacturing Project 4 (October): 76. https://doi.org/10.13140/RG.2.1.1849.4560.

**Dopfer, Kurt.** 2012. "The Origins of Meso Economics: Schumpeter's Legacy and beyond." Journal of Evolutionary Economics 22 (1): 133–160. https://doi.org/10.1007/s00191-011-0218-4.

**Dosi, Giovanni.** 1982. "Technological Paradigms and Technological Trajectories." Research Policy 11 (3): 147–62. https://doi.org/10.1016/0048-7333(82)90016-6

**Eslava, Marcela, Ana Ceclia Fieler, and Daniel Yi Xu.** 2015. "(Indirect) Input Linkages." American Economic Review 105 (5): 662–66. https://doi.org/10.1257/aer.p20151122.

**European Commission.** 2011. EU Industrial Structure 2011 Trends and Performance. Luxembourg: Publications Office of the European Union. https://doi.org/10.2769/28487.

**European Commission.** 2014. "EU Industrial Policy: The 2014 Communication for a European Industrial Renaissance." 2014. https://ec.europa.eu/growth/industry/policy/renaissance_en.

**Federal Ministry of Education and Research (BMBF).** 2010. "Ideas. Innovation. Prosperity. High-Tech Strategy 2020 for Germany." Innovation Policy

[33]
Framework Division. Bonn. [https://www.manufacturing-policy.eng.cam.ac.uk/documents-folder/policies/germany-ideas-innovation-prosperity-high-tech-strategy-2020-for-germany-bmbf/view](https://www.manufacturing-policy.eng.cam.ac.uk/documents-folder/policies/germany-ideas-innovation-prosperity-high-tech-strategy-2020-for-germany-bmbf/view).

**Floyd, Chris.** 1997. Managing Technology for Corporate Success. Gower Publishing, Ltd. [https://books.google.com/books?id=fSvcNJZ8nT4C](https://books.google.com/books?id=fSvcNJZ8nT4C).

**Hanusch, Horst, and Andreas Pyka, eds.** 2007. Elgar Companion to Neo-Schumpeterian Economics. Cheltenham: Edward Elgar Publishing Limited. [https://books.google.co.uk/books?id=P0zUrTYpMnkC](https://books.google.co.uk/books?id=P0zUrTYpMnkC).

**Hanusch, Horst, and Andreas Pyka.** 2007. "A Roadmap to Comprehensive Neo-Schumpeterian Economics," in: Elgar Companion to Neo-Schumpeterian Economics, chapter 70 Edward Elgar Publishing. [https://ideas.repec.org/h/elg/eechap/2973_70.html](https://ideas.repec.org/h/elg/eechap/2973_70.html)

**Hatzichronoglou, Thomas.** 1997. "Revision of the High-Technology Sector and Product Classification," OECD Science, Technology and Industry Working Papers. [https://doi.org/10.1787/134337307632](https://doi.org/10.1787/134337307632).

**Hausmann, Ricardo, and Dani Rodrik.** 2006. "Doomed To Choose: Industrial Policy as Predicament." America. [http://j.mp/2oWZa7W](http://j.mp/2oWZa7W).

**Inter-Secretariat Working Group on National Accounts.** 1993. System of National Accounts 1993. United Nations Statistical Commission. [https://unstats.un.org/unsd/nationalaccount/docs/1993sna.pdf](https://unstats.un.org/unsd/nationalaccount/docs/1993sna.pdf).

**Iran's Ministry of Industry Trade and Mine.** 2015. Industry, Mine, and Trade Strategic Plan. In Farsi. First Version. Tehran: Deputy of designing and planning. [http://en.mimt.gov.ir/general_content/222353-Industry-Mine-and-Trade-Strategic-Plan.html?t=General-content](http://en.mimt.gov.ir/general_content/222353-Industry-Mine-and-Trade-Strategic-Plan.html?t=General-content).

**Iran's Ministry of Industry Trade and Mine.** 2017. Industry, Mine, and Trade Strategic Plan. In Farsi. Second version. Tehran: Deputy of designing and planning.[http://mimt.gov.ir/parameters/mimt/modules/cdk/upload/content/file_manager/16526/BarnamehRahbordi.pdf](http://mimt.gov.ir/parameters/mimt/modules/cdk/upload/content/file_manager/16526/BarnamehRahbordi.pdf).

**Kee, Hiau Looi, and Heiwai Tang.** 2016. "Domestic Value Added in Exports : Theory and Firm Evidence from China." American Economic Review 106 (6): 1402–36. https://doi.org/10.1257/aer.20131687.

**Lall, Sanjaya, John Weiss, and Jinkang Zhang.** 2006. "The 'Sophistication' of Exports: A New Trade Measure." *World Development* 34 (2): 222–37. https://doi.org/10.1016/j.worlddev.2005.09.002.

**Leontief, Wassily.** 1971. "Theoretical Assumptions and Nonobserved Facts." The American Economic Review 61 (1): 1–7. http://www.jstor.org/stable/1910537.

**Leontief, Wassily.** 1976. "National Economic Planning : Methods and Problems." Challenge 19 (3): 6–11. http://www.jstor.org/stable/40719430.

**Leontief, Wassily.** 1986. Input-Output Economics. Second edi. New York: Oxford University Press. https://books.google.co.uk/books?id=hBDEXblq6HsC.

**List, Frederick.** 1856. National System of Political Economy. Philadelphia: J. B. LIPPINCOTT & CO. https://archive.org/details/in.ernet.dli.2015.213247.

**National Research Council.** 2014. Capturing Change in Science, Technology, and Innovation: Improving Indicators to Inform Policy. Washington, D.C.: National Academies Press. https://doi.org/10.17226/18606.

**Nelson, Richard R., and Sidney G. Winter.** 1983. "An Evolutionary Theory of Economic Change." The Economic Journal 93 (371): 652–54. https://doi.org/10.2307/2232409.

**Pavitt, Keith.** 1984. "Sectoral Patterns of Technical Change: Towards a Taxonomy and a Theory." *Research Policy* 13 (6): 343–73. https://doi.org/10.1016/0048-7333(84)90018-0.

**Peneder, Michael.** 2003. "Industry Classifications: Aim, Scope and Techniques." Journal of Industry, Competition and Trade 3 (1–2): 109–29. https://doi.org/10.1023/A:1025434721292.

**Peneder, Michael.** 2010. "Technological Regimes and the Variety of Innovation Behaviour: Creating Integrated Taxonomies of Firms and Sectors." Research

Policy 39 (3): 323–34. https://doi.org/10.1016/j.respol.2010.01.010.

**Perroux, François.** 1970. "Note on the Concept of Growth Poles." In Regional Economics: Theory and Practice, edited by David L. McKee, Robert D. Dean, and William H. Leahy, 93–104. New York: The Free Press. Translated from the 1955 original French, *Note sur la notion de pole de croissance*. https://archive.org/details/regionaleconomic00mckerich.

**Schumpeter, Joseph A.** 1923. Business-Cycles [Volume One, Chapter III]: A Theoretical, Historical, and Statistical Analysis of the Capitalist Process. First publ. New York: McGrow-Hill Book Company, Inc. (Reprinted 1939) https://archive.org/details/in.ernet.dli.2015.150123/.

**Schumpeter, Joseph A.** 1934. The Theory of Economic Development. Edited by Redvers Opie. Fifth prin. Cambrige: Harvard University Press. Translated from the 1911 original German, *Theorie der wirtschaftlichen Entwicklung*. https://archive.org/details/in.ernet.dli.2015.278364.

**Schumpeter, Joseph A.** 1989. Essays: On Entrepreneurs, Innovations, Business Cycles, and the Evolution of Capitalism. Edited by Richard Vernon Clemence. Eleventh. Classics in Economics. New Jersey: Transaction Publishers. Originally published in 1934 by Addison-wesley Press, Inc. https://books.google.com/books?id=WVMUGqMU5bAC.

**Soofi, Abdol.** 1992. "Industry Linkages, Indices of Variation and Structure of Production: An International Comparison." Economic Systems Research 4 (4): 349–76. https://doi.org/10.1080/09535319200000031.

**U.S. Department of Justice, and the Federal Trade Commission.** 2010. Horizontal Merger Guidelines. USA. https://www.justice.gov/sites/default/files/atr/legacy/2010/08/19/hmg-2010.pdf.

**Winter, Sidney G.** 1984. "Schumpeterian Competition in Alternative Technological Regimes." Journal of Economic Behavior & Organization 5 (3–4): 287–320. https://doi.org/10.1016/0167-2681(84)90004-0.

# APPENDICES

## FOR

## DESIGNING AN INDUSTRIAL POLICY FOR DEVELOPING COUNTRIES: A NEW APPROACH

**Appendix A**                                                                                   [For Online Publication]

**Measurement of Technology Content Added**
A model for the assessment of the Technology Content Added (TCA)

According to [Asian and Pacific Center for Transfer of Technology (APCTT)[17]] "The technology atlas team, 1987 p37: "The recognition of technology as an important strategic variable in development has led to the developing countries accepting the need for integrating technological considerations in the national socioeconomic planning process. However, one of the factors that has hampered these efforts appears to be the lack of suitable measures of technology".

"Technology can be considered to be the engine of growth for the national economy. Ordinarily technology is considered as something physical. Only rarely is it understood as a transformer of resources-not just the physical tools and facilities (hardware). In addition to the hardware, transformation of resources for economic growth requires human skills, accumulated knowledge, and institutional arrangements. The study presents a framework of the four basic components of technology for resources transformation, namely: 1) Technoware (object embodied technology); 2) Humanware (person embodied technology); 3) Infoware (document embodied technology); and 4) Orgaware (institution embodied technology)". (APTTC, 1986, p19)

"Technoware consists of tools, equipment, machines, vehicles, physical facilities, etc. Humanware refers to experiences, skills, knowledge, wisdom, creativity, etc. Inforware includes all kinds of documentation pertaining to process specifications, procedures, theories, observations, etc. Orgaware is required to facilitate the effective integration of Technoware, Humanware, and Inforware, and consists of management practices, linkages, etc." (Ibid, p22)

"If the pattern of the development of each of the four components of technology are examined, it is possible to perceive certain distinct phases in their growth process. These phases taken together may be called the Technology Life Chain and it is possible to describe a Life Chain for each component of technology." (Ibid, p29)

"The analysis of components of technology and the strength of life chain of each component give better insights for technology decision making. Such analysis can be applied to a variety of situations: assessment of technological capability in a specific field; assessment of national technological capability to generate technology; assessment of technological gap with respect to

---

[17] Asian and Pacific Center for Transfer of Technology (APCTT). 1987. "The Technology Atlas Project." Technological Forecasting and Social Change journal, Volume 32, Issue 1, p1-109. https://doi.org/10.1016/0040-1625(87)90003-5
https://doi.org/10.1016/0040-1625(87)90004-7        https://doi.org/10.1016/0040-1625(87)90005-9
https://doi.org/10.1016/0040-1625(87)90006-0        https://doi.org/10.1016/0040-1625(87)90007-2
https://doi.org/10.1016/0040-1625(87)90008-4

countries/industries/firms; assessment of technology content added in areas of relevance." (Ibid, p35)

"It has been proposed in earlier in this issue that the four components of technology, namely, Technoware, Humanware, Inforware, and Orgaware, are in fact the transformers of the inputs of a production system into outputs. Thus, any attempt to evaluate the transformation activity of a production system would have to necessarily examine the attributes of these four components." (Ibid, p38)

Economists have used the concept of value added to evaluate the monetary contribution of a transformation facility to the national economy. One definition of value added states that if the competitive condition that price equals unit costs is satisfied, the value added may be considered to be equal to the total cost of the factors of production used in the input-output transformation[18]. Since the four components of technology may be considered to be the equivalent of the factors of production, it may be useful to propose the measurement of the amount of "Technology Content" that is added at a transformation facility by these four components. (Ibid, p38)

Since the four components of technology, taken together, contribute towards the Technology Content of a transformation facility, a Technology Content Coefficient (TCC) may be defined by a multiplicative function as follows to describe a transformation facility:

$$TCC = \alpha \cdot T^{\beta_1} \cdot I^{\beta_2} \cdot H^{\beta_3} \cdot O^{\beta_4} \quad , \tag{A.1}$$

where the $\beta_i$'s may be called the intensity of contribution of each component towards the TCC and $\alpha$ is the "climate factor," which is an index of the country's commitment to technology as evaluated by the effectiveness with which technology activities are facilitated by the national environment. The multiplicative model is intuitively appealing due to the fact that it satisfies the properties listed below:

*Property 1*

T, I, H, 0 should all be strictly nonzero to ensure that TCC is nonzero. This is in accordance with the postulate that no transformation is possible without all four components of technology.

*Property 2*

Partial differentiation of TCC with respect to T results in the following expression:

$$\frac{\partial (TCC)}{\partial T} = \beta_1 \frac{(TCC)}{T} \tag{A.2}$$

Similar expressions can be obtained if partial differentiation is carried out with respect to H, I, and 0. Thus, if

$$0 < \beta_i < 1 \quad ,$$

then it meets the condition of the simultaneous requirement of all four components while satisfying the practically recognized phenomenon that the law of diminishing returns operates when attempts are made to increase technology levels by upgrading the level of only one component while keeping the others constant.

*Property 3*

The total differential of TCC may be expressed as follows:

$$d(TCC) = \frac{\partial (TCC)}{\partial T} \cdot dT + \frac{\partial (TCC)}{\partial I} \cdot dI + \frac{\partial (TCC)}{\partial H} \cdot dH + \frac{\partial (TCC)}{\partial O} \cdot dO \tag{A.3}$$

Thus,

$$d(TCC) = \beta_1 \frac{dT}{T} + \beta_2 \frac{dI}{I} + \beta_3 \frac{dH}{H} + \beta_4 \frac{dO}{O} \quad . \tag{A.4}$$

The proportionate increase of TCC would thus be equal to the sum of the proportionate increases of the four components weighted by the $\beta_i$'s.

*Property 4*

---

[18] Henderson, James Mitchell, and Richard E Quandt. 1980. Microeconomic Theory: A Mathematical Approach. Economics Handbook Series. McGraw-Hill. https://books.google.com/books?id=Fni7AAAAIAAJ.

If all the four components are increased by the same proportion $\mathcal{K}$, then eq. (A.3) would reduce to
$$\frac{d(TCC)}{TCC} = \mathcal{K}[\beta_1 + \beta_2 + \beta_3 + \beta_4] \, . \tag{A.5}$$
Thus, if
$$\beta_1 + \beta_2 + \beta_3 + \beta_4 < 1 \, , \tag{A.6}$$
then the TCC function obeys the condition of decreasing returns to scale. The operationalization of the multiplicative models, however, requires that estimates be made of T, I, H, 0, the $\beta_i$'s, and α. These estimation procedures are outlined next.

1. Estimation of T, I, H, 0: After an examination of the factors, using expert opinion, a score can be assigned for T, I, H, 0 on a range of 1-9. The highest value of 9 would refer to the best in the world, and all scoring would have to be done against this datum.

2. Estimation of the $\beta_i$'s: Property 3 shows that the proportionate increase of TCC is the sum of the proportionate increases of the four components weighted by the $\beta_i$'s. In any transformation facility, using expert opinion it should thus be possible to obtain estimates of the $\beta_i$'s by understanding the relative contributions that could result due to increases in the four components. However, the sum of the $\beta_i$'s should be less than unity according to Property 4. Well-established methods are available for arriving at such weightages using expert opinion[19].

3. Estimation of α: Any transformation facility can deliver its full technological capability only if the national technology climate is of a supportive nature. National level support may be implicit as well as explicit. At the firm level the extent of national support can be assessed by examining the effectiveness of relevant institutional services with respect to the functioning of the transformation facility. The maximum value of α = 1 while its minimum value would be 0.

Based on the above considerations, it would be possible to summarize the important attributes of the TCC computation as follows:

Expert opinion should be used to obtain estimates for the values of T, I, H, 0, a, and the $\beta_i$'s.
   The maximum value attainable by T, I, H, 0 is 9.
   The minimum value attainable by T, I, H, 0 is 1.
   The maximum value attainable by α is 1.
   The minimum value attainable by α is 0.
   $0 < \beta_i < 1$, i = 1, 2, 3, 4.
   $\beta_1 + \beta_2 + \beta_3 + \beta_4 < 1$.
   The theoretical maximum of TCC will be very close to 9.
   The theoretical minimum of TCC will be 0.
   The Technology Content Added (TCA) is thus defined as follows:
$$TCA = (TCC/9) \times EVA \, , \tag{A.7}$$
where EVA is the economic value added at the transformation facility. This implies that if the EVA has been obtained using the best technology (T, I, H, O), then the TCA is almost equal to the EVA. If not, the TCA is lower than the EVA. The value for EVA may be obtained quite easily from the management accounting system of a firm. (Ibid, p38-42)

---

[19] Eckenrode, Robert T. 1965. "Weighting Multiple Criteria." Management Science 12 (3): 180–92. https://doi.org/10.1287/mnsc.12.3.180.

**Appendix B** [For Online Publication]

**Technology classification in terms of the competitive impact**

Chris Floyd[20], 1997, by developing Arthur D. little's model[21], makes a classification of technologies in terms of their competitive impact. Technologies can be divided into four categories: base, key, pacing and emerging, indicating the scope of competitive advantage the technology offers and its level of maturity. Bellow, the definitions for these categories are followed:

**Base technology:** technologies that, although necessary and essential to practice well, offer little potential for competitive advantage. These technology are typically widespread and shared. Base technologies are commodity items which do not give significant competitive advantage but which are entry hurdles. Provided you have got them, you dot need to worry.

**Key technology:** technologies that are most critical to competitive success because they offer the opportunity for meaning full process or product differentiation. These technologies yield competitive advantage.

**Pacing technology:** technologies that have the potential to change the entire basis of competition but have not yet embodied in a product or process. These technologies often develop into key technologies. They are, may be, tomorrow's key technologies. They are technologies that are emerging from the R&D labs and beginning to be incorporated into niche products as a prelude to incorporate into the core product range if they prove successful. Well established players, strong in the base and key technologies can be caught out by other companies developing new substitute pacing technologies. It is very tempting to assume that your technology approach is the only viable one, and to fail to anticipate the threat of substitution posed by alternative technologies.

**Emerging technology:** technologies are those which may become tomorrow's pacing technologies. Still in the research stage, emerging technologies show promise, but are not guaranteed to become valuable.

the next step is to look at the maturity of the technologies, to identify those which are new and therefore of interest and those which are old, and therefore under potential threat. Building on the concepts developed earlier, classify the technologies as base, key, pacing or emerging, to identify those that have significant strategic impact. As discussed earlier, key and pacing technologies give a company competitive advantage. Emerging technologies could give competitive advantage in the future. Base technologies are commodities; necessary but conferring no advantage.

Technology maturity and strategic impact tend to go together. Emerging technologies are likely to be of the bottom of the 'S' curve. Pacing or key technologies tend to be moving up the 'S' curve, and those that are base tend to be mature and at the top of the 'S' curve (Figure B.1). But the correlation is not always perfect. Since it is direly related to competitive advantage, the strategic impact of a technology is industry specific. In contrast, technology maturity is not industry specific, as it is a measure of the evolution of a technology regardless of application. In practice, specific technologies are not readily transferable from the originating industry. So, although maturity is theoretically a measure independent of industry sector, it is normally determined by the sector that uses it most. Maturity can therefore be regarded as synonymous with strategic impact.

---

[20] Floyd, Chris. 1997. Managing Technology for Corporate Success. Gower Publishing, Ltd. https://books.google.com/books?id=fSvcNJZ8nT4C.

[21] Arthur D. Little, Inc., philip A Roussel, Kamal N Saad, and Tamara J Erickson. 1991. Third Generation R & D: Managing the Link to Corporate Strategy. Massachusetts: Harvard Business School Press. https://archive.org/details/thirdgenerationr00rous.

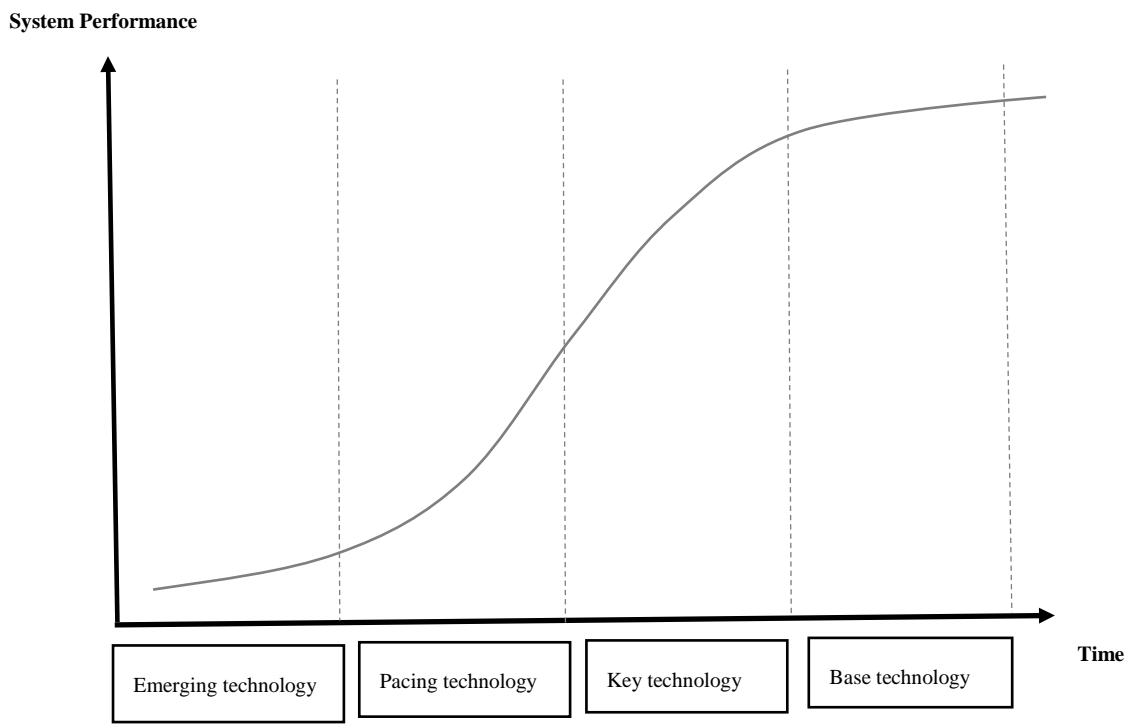

**FIGURE B.1-** THE TECHNOLOGY 'S' CURVE
**SOURCE:** ARTHUR D. LITTLE, INC. 1991.

# Appendix C [For Online Publication]

**Horizontal Merger Guidelines**[22]
"These Guidelines outline the principal analytical techniques, practices, and the enforcement policy of the Department of Justice and the Federal Trade Commission (the "Agencies") with respect to mergers and acquisitions involving actual or potential competitors ("horizontal mergers") under the federal antitrust laws." (U.S. Department of Justice; the Federal Trade Commission, 2010, p1)

**Market Concentration**
"Market concentration is often one useful indicator of likely competitive effects of a merger. In evaluating market concentration, the Agencies consider both the post-merger level of market concentration and the change in concentration resulting from a merger. Market shares may not fully reflect the competitive significance of firms in the market or the impact of a merger. They are used in conjunction with other evidence of competitive effects.

---

[22] U.S. Department of Justice, and the Federal Trade Commission. 2010. Horizontal Merger Guidelines. USA. https://www.justice.gov/sites/default/files/atr/legacy/2010/08/19/hmg-2010.pdf.

In analyzing mergers between an incumbent and a recent or potential entrant, to the extent the Agencies use the change in concentration to evaluate competitive effects, they will do so using projected market shares. A merger between an incumbent and a potential entrant can raise significant competitive concerns. The lessening of competition resulting from such a merger is more likely to be substantial, the larger is the market share of the incumbent, the greater is the competitive significance of the potential entrant, and the greater is the competitive threat posed by this potential entrant relative to others.

The Agencies give more weight to market concentration when market shares have been stable over time, especially in the face of historical changes in relative prices or costs. If a firm has retained its market share even after its price has increased relative to those of its rivals, that firm already faces limited competitive constraints, making it less likely that its remaining rivals will replace the competition lost if one of that firm's important rivals is eliminated due to a merger. By contrast, even a highly concentrated market can be very competitive if market shares fluctuate substantially over short periods of time in response to changes in competitive offerings. However, if competition by one of the merging firms has significantly contributed to these fluctuations, perhaps because it has acted as a maverick, the Agencies will consider whether the merger will enhance market power by combining that firm with one of its significant rivals.

The Agencies may measure market concentration using the number of significant competitors in the market. This measure is most useful when there is a gap in market share between significant competitors and smaller rivals or when it is difficult to measure revenues in the relevant market. The Agencies also may consider the combined market share of the merging firms as an indicator of the extent to which others in the market may not be able readily to replace competition between the merging firms that is lost through the merger.

The Agencies often calculate the Herfindahl-Hirschman Index ("HHI") of market concentration. The HHI is calculated by summing the squares of the individual firms' market shares, and thus gives proportionately greater weight to the larger market shares.[23]

$$HHI = \sum_{i=1}^{n} S_i^2 \qquad (C.1)$$

Where $S_i$ is the market share of firm, and $n$ is the number of firms, and market share of the each firm expressed as a whole number, not a decimal.

When using the HHI, the Agencies consider both the post-merger level of the HHI and the increase in the HHI resulting from the merger. The increase in the HHI is equal to twice the product of the market shares of the merging firms.[24]

Based on their experience, the Agencies generally classify markets into three types:
　　Unconcentrated Markets: HHI below 1500
　　Moderately Concentrated Markets: HHI between 1500 and 2500
　　Highly Concentrated Markets: HHI above 2500

The Agencies employ the following general standards for the relevant markets they have defined:
*Small Change in Concentration:* Mergers involving an increase in the HHI of less than 100 points are unlikely to have adverse competitive effects and ordinarily require no further analysis.

---

[23] For example, a market consisting of four firms with market shares of thirty percent, thirty percent, twenty percent, and twenty percent has an HHI of 2600 ($30^2 + 30^2 + 20^2 + 20^2 = 2600$). The HHI ranges from 10,000 (in the case of a pure monopoly) to a number approaching zero (in the case of an atomistic market). Although it is desirable to include all firms in the calculation, lack of information about firms with small shares is not critical because such firms do not affect the HHI significantly.

[24] For example, the merger of firms with shares of five percent and ten percent of the market would increase the HHI by 100 ($5 \times 10 \times 2 = 100$).

*Unconcentrated Markets:* Mergers resulting in unconcentrated markets are unlikely to have adverse competitive effects and ordinarily require no further analysis.

*Moderately Concentrated Markets:* Mergers resulting in moderately concentrated markets that involve an increase in the HHI of more than 100 points potentially raise significant competitive concerns and often warrant scrutiny.

*Highly Concentrated Markets:* Mergers resulting in highly concentrated markets that involve an increase in the HHI of between 100 points and 200 points potentially raise significant competitive concerns and often warrant scrutiny. Mergers resulting in highly concentrated markets that involve an increase in the HHI of more than 200 points will be presumed to be likely to enhance market power. The presumption may be rebutted by persuasive evidence showing that the merger is unlikely to enhance market power.

The purpose of these thresholds is not to provide a rigid screen to separate competitively benign mergers from anticompetitive ones, although high levels of concentration do raise concerns. Rather, they provide one way to identify some mergers unlikely to raise competitive concerns and some others for which it is particularly important to examine whether other competitive factors confirm, reinforce, or counteract the potentially harmful effects of increased concentration. The higher the post-merger HHI and the increase in the HHI, the greater are the Agencies' potential competitive concerns and the greater is the likelihood that the Agencies will request additional information to conduct their analysis." (U.S. Department of Justice; the Federal Trade Commission, 2010, p18)

## Appendix D [For Online Publication]

A tool for analyzing structural change is input-output analysis. Here, the focus is on interindustry transactions. The interindustry transactions or industry by industry flow table provides a summary of the industrial structure of an economy for a given year. It contains information on the values of flows of goods and services between industries and between different sectors of the economy. (Claus, 2009, p134)[25]

"Pioneer researchers in the field include Leontief[26] (1953) and Rasmussen[27] (1956). Leontief's work in this respect involved triangulation on the input-output table for the USA as a mechanism of understanding the internal structure of interindustry transactions. This analytical framework rested upon concepts of dependence, independence, hierarchy and circularity of industries. Rasmussen used an input-output model in measuring changes in the structure of production in Denmark between 1947 and 1949. In this seminal study, he proposed a method for measurement of industrial linkages using the open static input-output model." (Soofi, 1992)[28]

**Measurements of Backward and Forward Linkages**

---

[25] Claus, Iris. 2009. "New Zealand's Economic Reforms and Changes in Production Structure." Journal of Economic Policy Reform 12 (2): 133–43. https://doi.org/10.1080/17487870902872938.

[26] Leontief, Wassily. 1951. The Structure of American Economy, 1919-1939: An Empirical Application of Equilibrium Analysis. 2, reprint ed. the University of California: Oxford University Press. https://books.google.co.uk/books?id=8pEdAAAAIAAJ.

[27] Rasmussen, Poul Nørregaard. 1956. Studies in Inter-Sectoral Relations. Amsterdam, North-Holland: Einar Harcks Forlag. https://books.google.co.uk/books?id=ta-gAQAACAAJ.

[28] Rasmussen, Poul Nørregaard. 1956. Studies in Inter-Sectoral Relations. Amsterdam, North-Holland: Einar Harcks Forlag. https://books.google.co.uk/books?id=ta-gAQAACAAJ.

Hazari (1970)[29] explain Rasmussen's method: The gross output levels X's required to sustain a given vector of final demand F in the input-output model are determined by the following equation:

$$X = (I - A)^{-1} F \quad . \tag{D.1}$$

The analysis of the elements of the $(I - A)^{-1}$ would reveal the structure of the economy as well as that of the industry. Let us denote the elements of the $(I - A)^{-1}$ matrix by $(b_{ij})$'s. The sum of the column elements of the $(I - A)^{-1}$

$$\sum_{i=1}^{n} b_{ij} = b_{.j} \tag{D.2}$$

indicates the total input requirements for a unit increase in the final demand for the jth sector. In a similar way the sum of the row elements

$$\sum_{j=1}^{n} b_{ij} = b_{i.} \tag{D.3}$$

indicates the increase in the output of sector number i needed to cope with a unit increase in the final demand of all the industries. The averages

$$\frac{1}{n} b_{.j} \quad (j = 1, \ldots, m) \tag{D.4}$$

are interpreted by Rasmussen[30] ". . . as an estimate of the direct and indirect increase in output to be supplied by an industry chosen at random if the final demand for the products of industry number j (j = 1, ... , m) increases by one unit."

A similar interpretation has been given by Rasmussen to the set of averages

$$\frac{1}{n} b_{i.} \quad (i = 1, \ldots, m) \tag{D.5}$$

These indices are not suitable for making inter-industrial comparisons and for this purpose the set of averages in (4) and (5) are normalized by the overall average defined as

$$\frac{1}{n^2} \sum_{j=1}^{n} \sum_{i=1}^{n} b_{ij} = \frac{1}{n^2} \sum_{j=1}^{n} b_{.j} = \frac{1}{n^2} \sum_{i=1}^{n} b_{i.} \tag{D.6}$$

and thus we consider the indices

$$U_j = \frac{\frac{1}{n} b_{.j}}{\frac{1}{n^2} \sum_{j=1}^{n} b_{.j}} \tag{D.7}$$

and

$$U_i = \frac{\frac{1}{n} b_{i.}}{\frac{1}{n^2} \sum_{i=1}^{n} b_{i.}} \tag{D.8}$$

The indices $U_j$ and $U_i$ are termed by Rasmussen as the "Index of Power of Dispersion and Index of Sensitivity of Dispersion." $U_j$ and $U_i$ can also be interpreted as measures of Hirschman[31]'s backward and forward linkages.

Since the averages $1/n\, b_{.j}$ have been interpreted earlier showing the requirements of inputs if the final demand of industry number j increases by 1 unit, $U_j > 1$ then indicates that the industry draws heavily on the rest of the system, and vice versa, in case of $U_j < 1$. Similarly $U_i > 1$ indicates that the industry number i will have to increase its output more than others for a unit increase in final demand from the whole system.

The indices in equations (D.7) and (D.8) are based on the method of averaging. It is, how-ever, well known from the theory of statistics that averages are sensitive to extreme values and may give misleading results. Consequently -the indices in (7) and (D.8) do not fully describe the structure of a particular industry. To illustrate this it is possible that an increase in the final demand for the product of a particular industry characterized by a high index of power of dispersion may not affect other industries. Such a situation would arise if a particular industry draws heavily on one or a few industries only.

---

[29] Hazari, Bharat R. 1970. "Empirical Identification of Key Sectors in the Indian Economy." The Review of Economics and Statistics 52 (3): 301–5. http://www.jstor.org/stable/1926298.

[30] Rasmussen, 1956. Studies in Inter-Sectoral Relations, chap. 8, page 133.

[31] Hirschman, Albert O. 1958. The Strategy of Economic Development. The Strategy of Economic Development: V. 10. Yale University Press. https://books.google.com/books?id=wls-AAAAYAAJ.

In order to overcome this difficulty a measure of variability must be defined and the indices of coefficient of variation are defined as

$$V_j = \frac{\sqrt{\frac{1}{n-1}\sum_{i=1}^{n}(b_{ij}-\frac{1}{n}\sum_{i=1}^{n}b_{ij})^2}}{\frac{1}{n}\sum_{i=1}^{n}b_{ij}} \tag{D.9}$$

and

$$V_i = \frac{\sqrt{\frac{1}{n-1}\sum_{j=1}^{n}(b_{ij}-\frac{1}{n}\sum_{j=1}^{n}b_{ij})^2}}{\frac{1}{n}\sum_{j=1}^{n}b_{ij}} \tag{D.10}$$

A high Vj can be interpreted as showing that a particular industry draws heavily on one or a few sectors and a low Vj as an industry drawing evenly from the other sectors. The Vi's can be interpreted similarly.

A key sector can be defined as one in which (a) both Ui and Uj are greater than unity (Uj > 1, Ui> 1), and (b) both Vj and Vi are relatively low. One can easily interpret these in terms of Hirschman's terminology. Hirschman defines a key sector as one, which has a high forward as well as backward linkage. Since Uj and Ui have already been defined as backward and forward linkages it follows that any industry in which both Uj and Ui are greater than unity, can be defined as a key sector under Hirschman's definition. It should be noted that no restriction is stipulated in Hirschman's definition of the key sectors on the values of Vj and Vi and he thus disregards the "spread effects" of the development of an industry. These spread effects are exceedingly important from the point of view of industrial diversification and economic development.

All the following text is directly reflecting Soofi's work in 1992:

**Measures of Industry Interconnectedness**

Despite some important implications for interindustry economics, many researchers in this field make only passing references to the *V*s as measures of dispersion of interindustry flows. However, a close examination of the concept brings to the fore two important features of interindustry relationships: the significance of the number of direct and indirect industry connections and the importance of the magnitude of interindustry and intra-industry sales (purchases). (Soofi, 1992)

**A Measure of Concentration**

According to Soofi's study (1992): from equation (D.1) where $X = [X_1, …, X_n]'$ is the vector of gross output, *I* is the identity matrix, $A = [a_{ij}]$ is the matrix of technical coefficients, $a_{ij} > 0$, and $F = [F_1, … , F_n]'$ is the vector of final demand. Then for each $i_{th}$ sector,

$$X_i = \sum_{j=1}^{n} a_{ij} X_j + F_i \tag{D.11}$$

In this analysis, he initially concentrates on the intermediate sector by assuming that the $i_{th}$ sector's final demand delivery $F_i$ is equal to zero. This assumption, to be relaxed later, will allow us to normalize the elements of the matrix *A* with the corresponding row sums

$$\sum_{j=1}^{n} a_{ij} = a_{i.} \tag{D.12}$$

and column sums

$$\sum_{i=1}^{n} a_{ij} = a_{.j} \tag{D.13}$$

for all *i* and *j*.

Normalization of the rows of *A* results in an n x n matrix $C = [c_{ij}]$, with $c_{ij} = a_{ij}/a_{i.}$, $c_{ij} \geq 0$ and $\sum_{j=1}^{n} c_{ij} = 1$.

Complete uniformity of inter-sectoral distribution occurs when-all sectors receive the same quantity of input from the $i_{th}$ sector; hence, $c_{ij} = 1/n$ for all *j*. We have complete skewness in inter-sectoral distribution when only one sector receives the total output of the ith sector as input; therefore, $c_{ij} = 1$ for some j and $c_{ij}=0$ for all other *j*. Note that $V_i=0$ if and only if $c_{ij} = 1/n$ for all j= 1, ..., n, and $V_i=n-1$ if and only if $c_{ij} = 1$ for some *j* and $c_{ij} = 0$ for all other *j*. Therefore there is an inverse correspondence between the coefficient of variation and uniformity of inter-sectoral distribution. For example, consider the coefficient of variation of the $i_{th}$ row of A

$$V_i(a_{ij}) = \frac{\sqrt{1/n \sum_{j=1}^{n}(a_{ij}-1/n a_{.j})^2}}{1/n \, a_{.j}} \quad i = 1, \dots, n \tag{D.14}$$

Algebraic manipulation of equation (D.14) leads to

$$V_i(a_{ij}) = \sqrt{n}\sqrt{\sum_{j=1}^{n}(c_{ij} - \frac{1}{n})} \tag{D.15}$$

which implies

$$V_i^2(a_{ij}) = n\sum_{j=1}^{n}(c_{ij}^2 - 1) \tag{D.16}$$

Noting from equation (D.16) that $\max[V_i^2(a_{ij})] = n - 1$, define the following measure of concentration:

$$G_i(a_{ij}) = \sqrt{\max(V_i) - V_i} = \sqrt{n(1 - \sum_{j=1}^{n}c_{ij}^2)} \tag{D.17}$$

When there is no variation in a sector's sales to (purchases from) other sectors (when $c_{ij} = 1/n$ for all $j$), i.e. when $G = \sqrt{n-1}$, then the sum of industry sales (purchases) will also determine the number of direct sectoral ties. In this case complete uniformity in inter-sectoral transactions exists. Generally, however, given the sum of the $i_{th}$ industry's sales (purchases), a large value for G implies more direct industry ties. In contrast, a small measure of concentration (a small value for G) implies fewer interindustry sales or purchases. In the extreme case where $G = 0$ ($c_{ij} = 1$, for one $j$), total skewness in sectoral transactions prevails, which implies maximum concentration. Similarly, $G_j(a_{ij})$, $G_i^\omega(b_{ij})$ and $G_j^\omega(b_{ij})$ may be calculated.(ω: weighted)

Note that according to the foregoing analysis, in practice the ranking of $G$s should be in descending order of magnitude, which is congruous to the ranking of $U^\omega$s.

Along the lines of Diamond's work (1974)[32] we can construct a general index $GI$ representing the combined effects of $RU$ and $RG$, the ranks of $U$ and $G$ respectively, as follows[33]:

$$GI = \alpha RG + (1 - \alpha)RU \tag{D.18}$$

where $\alpha$ is the weight to be attached to the $G$ index. The parameter $\alpha$ reflects planners' preference for the sectors with uniform industry sales and purchases. This index generalizes Diamond's approach. It should eliminate the possibility of confusion arising from opposite ranking of the $U$s and $V$s.

Rewriting equation (D.18) as

$$GI = \alpha(RG - RU) + RU \tag{D.19}$$

we can observe the following possibilities. First, when $RG = RU$, then the ranking of $U$ or $G$ alone should suffice in decision-making. Second, for $RG > RU$, the sectors with a lower measure of concentration and high linkages are ranked lower than sectors with the same linkage value but a higher measure of concentration. Third, for $RG < RU$, the $GI$ value will lower the $U$ ranking of the sector. Accordingly, given two sectors with equal linkage index but different concentration measures, the $GI$ index will rank the sector with the larger concentration measure higher.

Note that the $GI$ index modifies the ranking of sectors with wide differences in values for $G$ and $U$. Also, the $GI$ index will have a small effect in the ranking of sectors with small differences between the $G$ and $U$ rankings.

**Entropy as a Measure of Variation**

---

[32] DIAMOND, J. 1974. "THE ANALYSIS OF STRUCTURAL CONSTRAINTS IN DEVELOPING ECONOMIES: A CASE STUDY*." Oxford Bulletin of Economics and Statistics 36 (2): 95–108. https://doi.org/10.1111/j.1468-0084.1974.mp36002002.x.

[33] I use U as a generic term to include all linkage indices regardless of the data used in their calculation.

From a review of the literature one can observe two parallel developments in the measurement of industry linkages and interconnectedness. The traditional approach, the multi-sectoral linkage method, emphasizes the quantitative importance (the output multipliers) of each sector in the economy. The number of direct and indirect industry ties is implicitly accounted for in these indices. The entropy-based and other holistic measures (including the index of diversification and indirect industry relatedness), however, tend to concentrate on measures of interconnectedness in an economy and provide a single scalar, a holistic measure, of the input-output table that summarizes the degree of interconnectedness of the table and is purportedly descriptive of the characteristics of the economy as a whole. (Soofi, 1992)

Soofi (1992) in his paper, instead of calculating a single holistic measure, calculates entropy-based sectoral measures of the dispersion of transactions in an input-output table. He uses the Shannon formula (Shannon, 1949)[34] to calculate the sectoral-holistic measures. These indices are then used to compare the degree of industry in interconnectedness and hence the structure of production of the economies under investigation.

To calculate the sectoral-holistic indices, transform the input coefficient matrix $A$ into the matrix of input coefficient proportions, $C$. These proportions are then the counterparts of the probabilities attached to the occurrence of n events. Hence they are subject to the same mathematical manipulations that allow the use of the Shannon formula. (Ibid)

The entropy $H_i$ of the $i_{th}$ sector is given by

$$H_i(a_{ij}) = \sum_{i=1}^{n} c_{ij} \log(\frac{1}{c_{ij}}) = -\frac{1}{a_{i.}}\sum_{j=1}^{n} a_{ij} \log a_{ij} + \log(\sum_{j=1}^{n} a_{ij}) \qquad (D.20)$$

Similarly the entropy of the $j_{th}$ sector is given by

$$H_j(a_{ij}) = \sum_{j=1}^{n} c_{ij} \log(\frac{1}{c_{ij}}) = -\frac{1}{a_{.j}}\sum_{i=1}^{n} a_{ij} \log a_{ij} + \log(\sum_{i=1}^{n} a_{ij}) \qquad (D.21)$$

Note that the $c_{ij}$ in equation (D.21) are defined as $c_{ij}= a_{ij}/a_{.j}$. Maximizing equation (D.20) subject to $\sum_{j=1}^{n} c_{ij} = 1$ and solving for $c_{ij}$ yields $c_{ij}= 1/n$, implying that the maximum entropy value is equal to $\log n$.

Accordingly, the entropy for each sector (row) is conceptually parallel to the coefficients of variation $V_j$; and the entropy for each column is a counterpart of the $V_i$. The row entropy for the $i_{th}$ sector, $H_i(a_{ij})$, is zero when the $j_{th}$ sector is the only sector which purchases additional output from' the $i_{th}$ sector after the $i_{th}$ sector delivers one dollar's worth of its output to the final demand. This is the minimum entropy sector. $H_i(a_{ij}) = \log n$ when all sectors of the, economy purchase an equal amount of output after the $i_{th}$ sector delivers one dollar's worth of its output to the final demand. This is the maximum entropy sector. The higher the variations in the sectoral response to a change in the delivery of the $i_{th}$ sector's output to the final demand, the lower the value of $H_i(a_{ij})$.

Similarly, the column entropy for the $j_{th}$ sector, $H_j(a_{ij})$, will be zero if the $j_{th}$ sector purchases additional output from only one industry in response to the $i_{th}$ sector's delivery of one dollar's worth of output to the final demand. $H_j(a_{ij}) = \log n$ if the $j_{th}$ sector uniformly increases its total intra-industry and interindustry purchases in response to a change in the $i_{th}$ sector's delivery of output to the final demand. The maximum/minimum entropy is used, then, in defining the interval for row/column entropy: $0 \leq H_i(a_{ij}) \leq \log^n$ and $0 \leq H_j(a_{ij}) \leq \log^n$ respectively.

To prepare a total requirement matrix for use in calculation of sectoral-holistic entropy indices, normalize the matrix by its row and column sums (the elements of the matrix $b_{ij}s$ are divided by the $\sum b_{ij} = b_{i.}$ and $b_{.j}$ respectively).

In addition to large numerous deliveries to the processing sector, and industry may also have economically significant deliveries to the final-demand sector of the economy. Therefore the economic importance of a sector is not exclusively determined by its deliveries to the intermediate sector of the economy. Additionally, one can cite examples of industries that exclusively deal with the final-demand sector of the economy. In such circumstances, however, the entropy measure as applied above will misrepresent the sector. To account appropriately for the sectors with strategic important final-demand deliveries, the entropy formula can be applied directly to the flow table.

---

[34] Shannon, Claude Elwood. 1948. "A Mathematical Theory of Communication." The Bell System Technical Journal 27 (3&4): 379-423,623-656. https://doi.org/10.1002/j.1538-7305.1948.tb01338.x.

To measure the impact of deliveries to the processing sectors as well as the final demand sectors, describe the economy by

$$X_i = a_{i1}X_1 + a_{i2}X_2 + \cdots + a_{in}X_n + F_i \quad i = 1, \ldots, n \tag{D.22}$$

To normalize the system of equations (D.22), divide both sides by $X_i$ and apply the entropy formula (D.20) to the proportions. The calculated entropy values measure interindustry sales as well as sectoral sales to the final demand.

The interpretation of the $H_i$ and $H_j$ that are based on (D.22) is straightforward. $H_i = 0$ when the $i_{th}$ sector sells to one sector only. $H_i = log\ (n + 1)$ when the $i_{th}$ sector sells an equal amount of output to all intermediate sectors as well as the final-demand sector of the economy. Also, when $H\ j = 0$ the $j_{th}$ sector buys from, one sector, and when $H_j = log\ n$ the $j_{th}$ sector purchases uniformly from all other sectors. Therefore, the entropy can measure the degree of industrial interconnectedness by measuring the dispersion of row and column elements in an input-output matrix.